\documentclass{ws-ijmpa}
\usepackage[super,compress]{cite}
\usepackage{graphicx}

\def\Journal#1#2#3#4{{#1} {\bf #2}, #3 (#4)}
\def\CPC{Chin. Phys. C}
\def\EPJC{Eur. Phys. J. C}
\def\IJMPA{Int. J. Mod. Phys. A}

\def\JCAP{J. Cosmol. Astropart. Phys.}

\def\JCAP{JCAP}
\def\JHEP{JHEP}

\def\JETPUSSR{JETP (USSR)}
 
\def\JPGNPP{J. Phys. G. Nucl. Part. Phys.}

\def\NPB{Nucl. Phys. B}

\def\PLB{Phys. Lett. B}

\def\PLBOLD{Phys. Lett.}

\def\PRL{Phys. Rev. Lett.}
\def\PRD{Phys. Rev. D}

\def\PTP{Prog. Theor. Phys.}

\def\ZETP{Zh. Eksp. Teor. Piz.}

\begin{document}
\markboth{Teruyuki Kitabayashi and Masaki Yasu\`{e}}{Seesaw model and two zero flavor neutrino texture}

\catchline{}{}{}{}{}
%
%%%%%%%%%%%%%%%%%%%%%%%%%%%%%%%%%%%%%%%%%%%%%%%%%%%%%%%%%%%%%%%%%%%%

\title{Seesaw model and two zero flavor neutrino texture}

\author{Teruyuki Kitabayashi and Masaki Yasu\`{e}}

\address{Department of Physics, Tokai University,\\
4-1-1 Kitakaname, Hiratsuka, Kanagawa 259-1292, Japan\\
teruyuki@tokai-u.jp, yasue@keyaki.cc.u-tokai.ac.jp}

\maketitle

\begin{history}
\received{Day Month Year}
\revised{Day Month Year}
\end{history}

%%-------------------------------------------------
%% Abstract
%%-------------------------------------------------
\begin{abstract}
In the two zero flavor neutrino mass matrix scheme with nonvanishing Majorana effective mass $M_{ee}$ for the neutrinoless double beta decay, four textures are compatible with observed data. We obtain the complete list of the possible textures of four zero Dirac neutrino mass matrix $m_D$ in the seesaw mechanism providing these four flavor neutrino textures. Explicit analytical analysis of $m_D$ turns out to provide the relation of $m_D \propto \sqrt{M_{ee}}$.
\end{abstract}

\ccode{PACS numbers:14.60.St:98.80.Cq}

%\tableofcontents

%% Main body
%%-------------------------------------------------

%%----------------------------------------------------------------------------------
\section{Introduction}
\label{sec:introduction}
%%----------------------------------------------------------------------------------
The seesaw mechanism provides a natural explanation of the smallness of the neutrino masses \cite{TypeISeesaw1,TypeISeesaw2,TypeISeesaw3,TypeISeesaw4,TypeISeesaw5,TypeISeesaw6}. The seesaw mechanism generates a $3 \times 3$ flavor neutrino mass matrix for light neutrinos constructed from a Dirac neutrino mass matrix $m_D$ and a heavy neutrino mass matrix $M_R$. One may analytically solve $m_D$ for a fixed $M_R$ in terms of elements of the flavor neutrino masses if some elements in $m_D$ are taken to vanish. For example, the Dirac mass matrices with four zero elements such as 
\begin{eqnarray}
m_D = 
\left(
  \begin{array}{ccc}
    0  & 0 & * \\
    *  & * & 0 \\
    *  & * & 0 \\
  \end{array}
\right) \ {\rm and} \
\left(
  \begin{array}{ccc}
    0  & * & * \\
    *  & 0 & 0 \\
    *  & 0 & *  \\
  \end{array}
\right),
\end{eqnarray}
are studied where the mark ``$*$" denotes a nonvanishing element \cite{fourZeroDirac1,fourZeroDirac2,fourZeroDirac3,fourZeroDirac4,fourZeroDirac5,fourZeroDirac6,fourZeroDirac7}. This type of matrix is called the four zero Dirac neutrino texture or four zero Yukawa texture.

Phenomenologically, $m_D$ is so adjusted to yield flavor neutrino masses that are consistent with observed data on neutrino oscillations. There have been various discussions of flavor neutrino mass matrix which is symmetric with respect to the flavor and induces to ensure the appearance of the observed neutrino mixings and masses \cite{ReviewOfMixingMatrix}. For example, flavor neutrino mass matrices, $M$, with two zeros such as  
\begin{eqnarray}
M = 
\left(
  \begin{array}{ccc}
    *  & * & 0   \\
    -  & 0 & *   \\
    -  & - & *  \\
  \end{array}
\right) \ {\rm and} \
\left(
  \begin{array}{ccc}
    *  & 0 & *   \\
    -  & *& *   \\
    -  & - & 0  \\
  \end{array}
\right),
\end{eqnarray}
are studied  where the mark ``$-$" denotes a symmetric partner \cite{twoZeroFlavor1,twoZeroFlavor2,twoZeroFlavor3,twoZeroFlavor4,twoZeroFlavor5}. This type of matrix is called the two zero flavor neutrino texture. If we require a nonvanishing Majorana effective mass $M_{ee}$ for the neutrinoless double beta decay \cite{doubleBetaDecay}, only four textures, which are classified as ${\rm B_1, B_2, B_3}$ and ${\rm B_4}$, are compatible with observed data in the two zero flavor neutrino mass matrix scheme \cite{twoZeroFlavor2}.

In this paper, we clarify the linkage between textures of the four zero Dirac neutrino mass matrix and two zero flavor neutrino textures of ${\rm B_1, B_2, B_3}$ and ${\rm B_4}$. We find that $m_D$ with four zero Dirac neutrino mass matrix is proportional to $\sqrt{M_{ee}}$, namely, $m_D$ rescaled with $\sqrt{M_{ee}}$, such as
\begin{eqnarray}
m_D \propto \sqrt{M_{ee}}
\left(
  \begin{array}{ccc}
    0  & 0 & 1 \\
    0  & * & * \\
    *  & * & 0 \\
  \end{array}
\right).
\end{eqnarray}
The usefulness of such rescaling with the factor $\sqrt{M_{ee}}$ has already been pointed out as a result of numerical calculations \cite{fourZeroDirac1,fourZeroDirac2,fourZeroDirac3,fourZeroDirac4,fourZeroDirac5,fourZeroDirac6,fourZeroDirac7}. Generally speaking, it is always possible to rescale $m_D$ with any factor.  However, rescaling $m_D$ with $\sqrt{M_{ee}}$ is physically meaningful because $M_{ee}$ is related to the neutrinoless double beta decay.  Furthermore, to manipulate this rescaling and to obtain systematic estimation of $m_D$ can be well provided by our method to evaluate flavor neutrino masses \cite{twoZeroFlavor5,KitabayashiYasue2016}. The complete list of the explicit analytical expressions of the rescaled mass matrix with $\sqrt{M_{ee}}$ is useful to see how the effect of the neutrinoless double beta decay controls the Dirac neutrino masses. This complete list is shown by analytically for the first time in this paper.

In Sec.\ref{sec:brief_review}, we show a brief review of the seesaw model and  two zero flavor neutrino texture. In Sec.\ref{sec:Dirac_neutrino_mass_matrix}, we discuss how to obtain results indicating $m_D \propto \sqrt{M_{ee}}$ to derive explicit analytical expressions instead of numerical calculations and list the possible textures of the four zero Dirac neutrino mass matrix compatible with the two zero flavor neutrino textures of ${\rm B_1, B_2, B_3}$ and ${\rm B_4}$. The final section Sec.\ref{sec:summary} is devoted to summary and discussions, which include a simple argument on the creation of the baryon number of the universe via leptogenesis to see the usefulness of our analytical expressions of $m_D$ rescaled with $\sqrt{M_{ee}}$.

%%----------------------------------------------------------------------------------
\section{\label{sec:brief_review} Brief review}
%%----------------------------------------------------------------------------------
%%----------------------------------------------------------------------------------
\subsection{Seesaw model}
%%----------------------------------------------------------------------------------
The neutrino mass terms in the seesaw mechanism can be written as 
\begin{eqnarray}
\mathcal{L}_m= - \overline{\nu}_Lm_DN_R - \frac{1}{2}\overline{N}_R^c M_RN_R + h.c.,
\end{eqnarray}
where $\nu_L = (\nu_e,\nu_\mu,\nu_\tau)^T$ and $N_R=(N_1,N_2,N_3)^T$ denote the left-handed (light) neutrinos and the right-handed (heavy) neutrinos, respectively. We assume that the mass matrix of the heavy neutrinos $M_R=\textrm{diag}.(M_1,M_2,M_3)$ is diagonal and real. The Dirac neutrino mass matrix $m_D$ is a $3 \times 3$ square matrix, which can be written by 9 parameters ($a_1, a_2, a_3, b_1, b_2, b_3, c_1, c_2, c_3$) with fixed $M_R$ as \cite{Chang2004PLB}
\begin{eqnarray}
m_D = 
\left(
  \begin{array}{ccc}
    \sqrt{M_1}a_1  & \sqrt{M_2}b_1 & \sqrt{M_3}c_1  \\
    \sqrt{M_1}a_2  & \sqrt{M_2}b_2 & \sqrt{M_3}c_2 \\
    \sqrt{M_1}a_3  & \sqrt{M_2}b_3 & \sqrt{M_3}c_3 \\
  \end{array}
\right).
\label{Eq:mD}
\end{eqnarray}
We obtain the symmetric $3 \times 3$ flavor neutrino mass matrix 
\begin{eqnarray}
M=\left(
  \begin{array}{ccc}
    M_{ee} & M_{e\mu} & M_{e\tau} \\
     -      & M_{\mu\mu} & M_{\mu\tau} \\
     -      &  -      & M_{\tau\tau} \\
  \end{array}
\right),
\label{Eq:Mnu}
\end{eqnarray}
where
\begin{eqnarray}
&& M_{ee} = a_1^2 + b_1^2+c_1^2, \  M_{e\mu} = a_1a_2 + b_1b_2 + c_1c_2, \nonumber \\ 
&& M_{e\tau} = a_1a_3 + b_1b_3 + c_1c_3,\ M_{\mu\mu} = a_2^2 + b_2^2 + c_2^2, \nonumber \\
&& M_{\mu\tau} = a_2a_3 + b_2b_3 +c_2c_3,\ M_{\tau\tau} = a_3^2 + b_3^2  +c_3^2.
\end{eqnarray}
% 

%%----------------------------------------------------------------------------------
\subsection{Two zero flavor neutrino texture}
%%----------------------------------------------------------------------------------
The flavor neutrino mass matrix $M$ is related to neutrino masses $m_1,m_2,m_3$ obtained from the diagonalizations of $M$ using the unitary matrix $U$: ${\rm diag.}(m_1, m_2, m_3)= U^T M U$. We adopt the following parameterization of $U$ \cite{PMNS1,PMNS2,PDG}
\begin{eqnarray}
U =
\left( 
\begin{array}{*{20}{c}}
1 & 0 & 0 \\
0 & c_{23} & s_{23} \\
0 & -s_{23} & c_{23} \\
\end{array}
\right)
\left( 
\begin{array}{*{20}{c}}
c_{13} & 0 & s_{13}e^{-i\delta} \\
0 & 1 & 0 \\
-s_{13}e^{i\delta} & 0 & c_{13} \\
\end{array}
\right) 
\left( 
\begin{array}{*{20}{c}}
c_{12} & s_{12} & 0 \\
-s_{12} & c_{12} & 0 \\
0 & 0 & 1 \\
\end{array}
\right)
\left( 
\begin{array}{*{20}{c}}
e^{i\phi_1/2}& 0 & 0 \\
0 & e^{i\phi_2/2}& 0 \\
0 & 0 & e^{i\phi_3/2} \\
\end{array}
\right), 
\label{Eq:U_0}
\end{eqnarray}
where $c_{ij}=\cos\theta_{ij}$ and $s_{ij}=\sin\theta_{ij}$ (as well as $t_{ij}=\tan\theta_{ij}$) and $\theta_{ij}$ represents the $\nu_i$-$\nu_j$ mixing angle ($i,j$=1,2,3). The phases $\delta$ and $\phi_{1,2,3}$ stand for one CP-violating Dirac phase and three Majorana phases, respectively, where CP-violating Majorana phases are specified by two combinations made of $\phi_{1,2,3}$ such as $\alpha_i = \phi_i - \phi_1$ ($i=2,3$) in place of $\phi_{1,2,3}$ giving $K^{PDG} = {\rm diag.}(1, e^{i\alpha_2/2}, e^{i\alpha_3/2})$ \cite{CPViolationOrg1,CPViolationOrg2,CPViolationOrg3}.

In the two zero flavor neutrino mass matrix scheme, there are 15 possible combinations of two vanishing independent elements in the flavor neutrino mass matrix. If we require a nonvanishing Majorana effective mass $M_{ee}$ for the neutrinoless double beta decay, the viable textures are the following \cite{twoZeroFlavor1,twoZeroFlavor2,twoZeroFlavor3,twoZeroFlavor4,twoZeroFlavor5}
\begin{eqnarray}
&& B_1:
\left( 
\begin{array}{*{20}{c}}
M_{ee} & M_{e\mu} & 0 \\
- & 0 & M_{\mu\tau} \\
- & - & M_{\tau\tau} \\
\end{array}
\right),
\
B_2:
\left( 
\begin{array}{*{20}{c}}
M_{ee} & 0 & M_{e\tau} \\
-& M_{\mu\mu} & M_{\mu\tau} \\
- & - & 0 \\
\end{array}
\right),
\nonumber \\
&&B_3:
\left( 
\begin{array}{*{20}{c}}
M_{ee} & 0 & M_{e\tau} \\
- & 0 & M_{\mu\tau} \\
- & - & M_{\tau\tau} \\
\end{array}
\right),
\
B_4:
\left( 
\begin{array}{*{20}{c}}
M_{ee} & M_{e\mu} & 0 \\
- & M_{\mu\mu} & M_{\mu\tau} \\
- & - & 0 \\
\end{array}
\right),
\nonumber \\
&&C:
\left( 
\begin{array}{*{20}{c}}
M_{ee} & M_{\mu\tau} & M_{e\tau} \\
- & 0 & M_{\mu\tau} \\
- & - & 0 \\
\end{array}
\right).
\label{Eq:B1B2B3B4C}
\end{eqnarray}
On the other hand, according to Dev et.al., \cite{twoZeroFlavor2} as well as from recent T2K, Super-Kamiokande and NO$\nu$A results, it is suggested that the ${\rm B_1}$, ${\rm B_2}$, ${\rm B_3}$ and ${\rm B_4}$ textures among five patterns in Eq.(\ref{Eq:B1B2B3B4C}) are compatible with the latest neutrino oscillation data \cite{twoZeroFlavor4}. In this paper, we restrict our attention to the ${\rm B_1,B_2,B_3}$ and ${\rm B_4}$ textures.

Our recent discussions \cite{KitabayashiYasue2016} have found that the flavor neutrino masses and mass eigenstates for ${\rm B_1,B_2,B_3}$ and ${\rm B_4}$ depend on $M_{ee}$ (with mixing angles and phases fixed). For textures labelled by $X={\rm B_1,B_2,B_3,B_4}$, we have  
\begin{eqnarray}
M_a &=& f_a^X(\theta_{12},\theta_{23}, \theta_{13},\delta)M_{ee}, \nonumber \\
\tilde{m}_j &\equiv& m_je^{-i\phi_j} = f_j^X(\theta_{12},\theta_{23}, \theta_{13},\delta)M_{ee}, 
\label{Eq:m_f_Mee}
\end{eqnarray}
where $a=e\mu,\mu\tau, \tau\tau, \cdots$ and $j=1,2,3$.  

More concretely, we obtain the following expressions for the ${\rm B_1}$ texture:
\begin{eqnarray}
&& M_{e\mu}=f_{e\mu}^{\rm B1}M_{ee}, \
M_{\mu\tau}=f_{\mu\tau}^{\rm B1} M_{ee}, \
M_{\tau\tau}=f_{\tau\tau}^{\rm B1} M_{ee},
\label{Eq:M_B1} \\
&&\tilde{m}_1 =f_1^{\rm B1} M_{ee}, \
\tilde{m}_2 =f_2^{\rm B1} M_{ee}, \
\tilde{m}_3 =f_3^{\rm B1} M_{ee},
\label{Eq:m_B1}
\end{eqnarray}
where
\begin{eqnarray}
f_{e\mu}^{\rm B1}&=&-\frac{A_1}{c_{23}B_1 + s_{23}C_1},
\nonumber \\
f_{\mu\tau}^{\rm B1}&=&-A_1\frac{-c_{23}B_3 + s_{23}C_3}{c_{23}B_1 + s_{23}C_1} - \frac{1-e^{-2i\delta}}{2} \sin 2\theta_{23},
\nonumber \\
f_{\tau\tau}^{\rm B1}&=&-A_1\frac{c_{23}B_2+ s_{23}C_2}{c_{23}B_1 + s_{23}C_1} + A_2,
\label{Eq:f_a_B1}
\end{eqnarray}
and
\begin{eqnarray}
f_1^{\rm B1}&=&A_1\frac{ \frac{t_{12}c_{23}}{c_{13}} +t_{13}s_{23}e^{i\delta} }{c_{23}B_1 + s_{23}C_1} + 1,
\nonumber \\
f_2^{\rm B1}&=&A_1\frac{ -\frac{c_{23}}{c_{13}t_{12}} +t_{13}s_{23}e^{i\delta} }{c_{23}B_1 + s_{23}C_1} + 1,
\nonumber \\
f_3^{\rm B1}&=&A_1\frac{ -\frac{s_{23}}{t_{13}}e^{-i\delta} }{c_{23}B_1 + s_{23}C_1} + e^{-2i\delta},
\label{Eq:f_j_B1}
\end{eqnarray}
with 
\begin{eqnarray}
A_1&=&c_{23}^2+s_{23}^2 e^{-2i\delta}, \nonumber \\
A_2&=&s_{23}^2+c_{23}^2 e^{-2i\delta}, \nonumber \\
B_1&=& \frac{2c_{23}^2}{c_{13}\tan 2\theta_{12}} - t_{13}\sin 2\theta_{23} e^{-i\delta}, \nonumber \\
B_2&=& \frac{2s_{23}^2}{c_{13}\tan 2\theta_{12}} + t_{13}\sin 2\theta_{23} e^{-i\delta}, \nonumber \\
B_3&=&  \frac{\sin 2\theta_{23}}{c_{13}\tan 2\theta_{12}} +t_{13} \cos 2\theta_{23} e^{-i\delta}, \nonumber \\
C_1&=& 2  \left( \frac{s_{23}^2 e^{-i\delta}}{\tan 2\theta_{13}} -\frac{t_{13} c_{23}^2 e^{i\delta}}{2} \right), \nonumber \\
C_2&=& 2  \left( \frac{c_{23}^2 e^{-i\delta}}{\tan 2\theta_{13}} -\frac{t_{13} s_{23}^2 e^{i\delta}}{2} \right), \nonumber \\
C_3&=&  \sin 2\theta_{23} \left( \frac{e^{-i\delta}}{\tan 2\theta_{13}} +\frac{t_{13} e^{i\delta}}{2} \right).
\end{eqnarray}

Similarly, for the ${\rm B_2}$ texture, we obtain the functions of $f_a^X$ and $f_j^X$ as
\begin{eqnarray}
f_{e\tau}^{\rm B2}&=&\frac{A_2}{s_{23}B_2 - c_{23}C_2},
\nonumber \\
f_{\mu\mu}^{\rm B2}&=&A_2\frac{-s_{23}B_1+ c_{23}C_1}{s_{23}B_2 - c_{23}C_2} + A_1,
\\
f_{\mu\tau}^{\rm B2}&=&A_2\frac{s_{23}B_3 + c_{23}C_3 }{s_{23}B_2 - c_{23}C_2} - \frac{1-e^{-2i\delta}}{2} \sin 2\theta_{23},
\nonumber 
\label{Eq:f_a_B2}
\end{eqnarray}
\begin{eqnarray}
f_1^{\rm B2}&=&A_2\frac{ \frac{t_{12}s_{23}}{c_{13}} -t_{13}c_{23}e^{i\delta} }{s_{23}B_2 - c_{23}C_2} + 1,
\nonumber \\
f_2^{\rm B2}&=&A_2\frac{ -\frac{s_{23}}{c_{13}t_{12}} -t_{13}c_{23}e^{i\delta} }{s_{23}B_2 - c_{23}C_2} + 1,
\nonumber \\
f_3^{\rm B2}&=&A_2\frac{ \frac{c_{23}}{t_{13}}e^{-i\delta} }{s_{23}B_2 - c_{23}C_2} + e^{-2i\delta},
\label{Eq:f_j_B2}
\end{eqnarray}
for the ${\rm B_3}$ texture, 
\begin{eqnarray}
f_{e\tau}^{\rm B3}&=&\frac{A_1}{s_{23}B_1 - c_{23}C_1},
\nonumber \\
f_{\mu\tau}^{\rm B3} &=&A_1\frac{s_{23}B_3 + c_{23}C_3 }{s_{23}B_1 - c_{23}C_1} - \frac{1-e^{-2i\delta}}{2} \sin 2\theta_{23},
\nonumber \\
f_{\tau\tau}^{\rm B3}&=&2A_1\frac{(s_{23}B_3+ c_{23}C_3)\cos 2\theta_{23} - s_{23}t_{13}e^{-i\delta}}{\sin 2\theta_{23}(s_{23}B_1 - c_{23}C_1)} 
- \frac{1-e^{-2i\delta}}{2} \cos 2\theta_{23},
\label{Eq:f_a_B3}
\end{eqnarray}
\begin{eqnarray}
f_{1}^{\rm B3}&=&A_1\frac{ \frac{t_{12}s_{23}}{c_{13}} -t_{13}c_{23}e^{i\delta} }{s_{23}B_1 - c_{23}C_1} + 1,
\nonumber \\
f_{2}^{\rm B3}&=&A_1\frac{ -\frac{s_{23}}{c_{13}t_{12}} -t_{13}c_{23}e^{i\delta} }{s_{23}B_1 - c_{23}C_1} + 1,
\nonumber \\
f_{3}^{\rm B3}&=&A_1\frac{ \frac{c_{23}}{t_{13}}e^{-i\delta} }{s_{23}B_1 - c_{23}C_1} + e^{-2i\delta},
\label{Eq:f_j_B3}
\end{eqnarray}
and, for the ${\rm B_4}$ texture,
\begin{eqnarray}
f_{e\mu}^{\rm B4}&=&-\frac{A_2}{c_{23}B_2 + s_{23}C_2},
\nonumber \\
f_{\mu\mu}^{\rm B4}&=&-A_2\frac{c_{23}B_1+ s_{23}C_1}{c_{23}B_2 + s_{23}C_2} + A_1,
\\
f_{\mu\tau}^{\rm B4}&=&-A_2\frac{-c_{23}B_3 + s_{23}C_3 }{c_{23}B_2 + s_{23}C_2} - \frac{1-e^{-2i\delta}}{2} \sin 2\theta_{23}, \nonumber
\label{Eq:f_a_B4}
\end{eqnarray}
\begin{eqnarray}
f_1^{\rm B4}&=&A_2\frac{ \frac{t_{12}c_{23}}{c_{13}} + t_{13}s_{23}e^{i\delta} }{c_{23}B_2 + s_{23}C_2} + 1,
\nonumber \\
f_2^{\rm B4}&=&-A_2\frac{ \frac{c_{23}}{c_{13}t_{12}} -t_{13}s_{23}e^{i\delta} }{c_{23}B_2 + s_{23}C_2} + 1,
\nonumber \\
f_3^{\rm B4}&=&-A_2\frac{ \frac{s_{23}}{t_{13}}e^{-i\delta} }{c_{23}B_2 + s_{23}C_2} + e^{-2i\delta}.
\label{Eq:f_j_B4}
\end{eqnarray}
%

%%----------------------------------------------------------------------------------
\section{\label{sec:Dirac_neutrino_mass_matrix} Dirac neutrino mass matrix}
%%----------------------------------------------------------------------------------
%%----------------------------------------------------------------------------------
\subsection{Allowed combinations of four zeros}
%%----------------------------------------------------------------------------------
For the sake of simplicity, we omit $\sqrt{M_1}, \sqrt{M_2}$ and $\sqrt{M_3}$ in Eq.(\ref{Eq:mD}) and use $\tilde{m}_D$ defined by 
\begin{eqnarray}
\tilde{m}_D = 
\left(
  \begin{array}{ccc}
    a_1  & b_1 & c_1  \\
    a_2  & b_2 & c_2 \\
    a_3  & b_3 & c_3 \\
  \end{array}
\right).
\end{eqnarray}
Because the nearly degenerated mass pattern $|m_1| \sim |m_2| \sim |m_3|$ is compatible with observed data for the two zero flavor textures \cite{twoZeroFlavor1,twoZeroFlavor2,twoZeroFlavor3,twoZeroFlavor4,twoZeroFlavor5}, we require $rank(m_D)=rank(\tilde{m}_D)=3$ for the Dirac mass matrix: e.g., if we allow $rank(m_D)<3$, at least one of the neutrino masses $m_1,m_2,m_3$ should vanish and we cannot guarantee the nearly degenerate mass pattern. The Dirac mass matrix with all zero entries in a row or a column such as 
\begin{eqnarray}
\left(
  \begin{array}{ccc}
    0  & b_1 & c_1  \\
    0  & b_2 & c_2 \\
    0  & b_3 & c_3 \\
  \end{array}
\right) \ {\rm and} \
\left(
  \begin{array}{ccc}
    0  & 0 & 0  \\
    a_2  & b_2 & c_2 \\
    a_3  & b_3 & c_3 \\
  \end{array}
\right),
\end{eqnarray}
and with same row or same column such as
\begin{eqnarray}
\left(
  \begin{array}{ccc}
    a_1  & a_1 & c_1  \\
    a_2  & a_2 & c_2 \\
    a_3  & a_3 & c_3 \\
  \end{array}
\right) \ {\rm and} \
\left(
  \begin{array}{ccc}
    a_1  & b_1 & c_1  \\
    a_1  & b_1 & c_1 \\
    a_3  & b_3 & c_3 \\
  \end{array}
\right),
\end{eqnarray}
are excluded.

For the ${\rm B_1}$ texture, where $M_{\mu\mu}=0$, if two elements in $(a_2,b_2,c_2)$ are zero, such as $a_2=b_2=0$, the remaining element should be also zero for $M_{\mu\mu}=a_2^2+b_2^2+c_2^2=0$; however, $a_2=b_2=c_2=0$ yields $rank(m_D)<3$. The possible textures take the following four cases for $\{a_2,b_2,c_2\}$: 
\begin{itemize}
\item $a_2=0$ and $c_2=i\sigma b_2 \neq 0$,
\item $b_2=0$ and $c_2=i\sigma a_2 \neq 0$,
\item $c_2=0$ and $b_2=i\sigma a_2 \neq 0$,
\item $a_2 \neq b_2 \neq c_2 \neq 0$,
\end{itemize}
where $\sigma = \pm 1$.

In the case of $a_2=0$, the Dirac mass matrix should be 
\begin{eqnarray}
\left(
  \begin{array}{ccc}
    a_1  & b_1 & c_1  \\
    0  & b_2 & i\sigma b_2 \\
    a_3  & b_3 & c_3 \\
  \end{array}
\right),
\end{eqnarray}
with additional three zero elements in the 1st and 3rd rows. The number of combinations of additional three zero entries from six entries $\{a_1,b_1,c_1,a_3,b_3,c_3\}$ is $_6C_3=20$. The six combinations: 
\begin{eqnarray}
(a_1,a_3,b_1), \ (a_1,a_3,b_3), \ (a_1,a_3,c_1), \ (a_1,a_3,c_3), (a_1,b_1,c_1), \ (a_3,b_3,c_3), 
\end{eqnarray}
yield $rank(m_D)<3$, three combinations: 
\begin{eqnarray}
(a_3,b_1,c_1), \ (b_1,b_3,c_1), \ (b_1,c_1,c_3),
\end{eqnarray}
yield $M_{e\mu}=0$, three combinations: 
\begin{eqnarray}
(a_1,b_3,c_3), \ (b_1,b_3,c_3), \ (b_3,c_1,c_3), 
\end{eqnarray}
yield $M_{\mu\tau}=0$, four combinations:
\begin{eqnarray}
&& (a_1,b_1,b_3), \ (a_1,c_1,c_3), \ (a_3,b_1,b_3), \ (a_3,c_1,c_3), 
\end{eqnarray}
yield texture one zero. These 16 combinations should be excluded. The following four combinations
\begin{eqnarray}
&& {\rm B_1}(a_1,a_2,b_1,c_3), \ {\rm B_1}(a_1,a_2,b_3,c_1), \nonumber \\ 
&& {\rm B_1}(a_2,a_3,b_1,c_3), \ {\rm B_1}(a_2,a_3,b_3,c_1), 
\end{eqnarray}
are allowed in the case of $a_2=0$ for the ${\rm B_1}$ texture, where ${\rm B_1}(a_1,a_2,b_1,c_3)$ means $a_1=a_2=b_1=c_3=0$. From similar discussions, the allowed combinations in the case of $b_2=0$, $c_2=0$ and $a_2 \neq b_2 \neq c_2$ for the ${\rm B_1}$ texture are obtained. The complete list of the allowed combinations for the ${\rm B_1}$ as well as ${\rm B_2, B_3}$ and ${\rm B_4}$ will be shown in the next subsection. 

We note that the maximum number of zero entries in the Dirac mass matrix compatible with ${\rm B_1,B_2,B_3}$ and ${\rm B_4}$ is four. For example, in the ${\rm B_1}(a_1,a_2,b_1,c_3)$ case, more number of zeros yields $rank(m_D)<3$ or $M_{\mu\tau}=0$.

%%----------------------------------------------------------------------------------
\subsection{List of the four zero Dirac mass matrix}
%%----------------------------------------------------------------------------------
For the case of ${\rm B_1}(a_1,a_2,b_1,c_3)$, the four zero Dirac mass matrix 
\begin{eqnarray}
\tilde{m}_D &=& \left(
  \begin{array}{ccc}
    0  & 0 & c_1  \\
    0  & b_2 & i\sigma b_2 \\
    a_3  & b_3 & 0 \\
  \end{array}
\right),
\end{eqnarray}
yields two zero flavor neutrino mass matrix as 
\begin{eqnarray}
M&=&\left(
  \begin{array}{ccc}
    c_1^2  & i\sigma b_2c_1 & 0  \\
    -  & 0 & b_2b_3 \\
    -  & - & a_3^2+b_3^2 \\
  \end{array}
\right).
\end{eqnarray}
From Eq.(\ref{Eq:M_B1}), we have the following relations 
\begin{eqnarray}
M_{ee}&=&c_1^2, \ M_{e\mu}=i\sigma c_1b_2=f_{e\mu}^{\rm B1}M_{ee}, \\
M_{\mu\tau}&=&b_2 b_3=f_{\mu\tau}^{\rm B1} M_{ee},  \
M_{\tau\tau}=a_3^2+b_3^2=f_{\tau\tau}^{\rm B1} M_{ee},\nonumber
\end{eqnarray}
and the nonvanishing entries in the Dirac mass matrix are calculated as 
\begin{eqnarray}
a_3 &=& \pm \sqrt{M_{ee}} \sqrt{f_{\tau\tau}^{\rm B1} + (f_{\mu\tau}^{\rm B1}/f_{e\mu}^{\rm B1})^2},\nonumber \\
b_2 &=& \mp i\sigma \sqrt{M_{ee}} f_{e\mu}^{\rm B1}, \
b_3=\pm i\sigma \sqrt{M_{ee}} (f_{\mu\tau}^{\rm B1}/f_{e\mu}^{\rm B1}), \nonumber \\
c_1 &=& \pm \sqrt{M_{ee}}.
\end{eqnarray}
As we have already mentioned in the introduction, a four zero Dirac neutrino mass matrix consistent with the ${\rm B_1}$ texture is proportional to $\sqrt{M_{ee}}$.

The complete list of the four zero Dirac mass matrices $m_D$ compatible ${\rm B_1, B_2, B_3}$ and ${\rm B_4}$ are obtained as follows. We show $\tilde{m}_D$ instead of $m_D$.

For the ${\rm B_1}$ texture with $f^{\rm B1} =\sqrt{f_{\tau\tau}^{\rm B1} + \left(\frac{f_{\mu\tau}^{\rm B1}}{f_{e\mu}^{\rm B1}}\right)^2}$, there are three variations:
\begin{description}
\item[1) ${\rm B_1}(a_1,a_2,b_1,c_3)$]
\begin{eqnarray}
\tilde{m}_D=\pm\sqrt{M_{ee}}
\left(
  \begin{array}{ccc}
    0  & 0 & 1  \\
    0  & - i\sigma f_{e\mu}^{\rm B1} & f_{e\mu}^{\rm B1}\\
    f^{\rm B1} & \frac{i \sigma f_{\mu\tau}^{\rm B1}}{f_{e\mu}^{\rm B1}} & 0 \\
  \end{array}
\right),
\end{eqnarray}
and
\begin{description}
\item[${\rm B_1}(a_1,a_2,b_3,c_1)$] $b_i \rightarrow -c_i$, $c_i \rightarrow b_i$,
\item[${\rm B_1}(a_1,b_1,b_2,c_3)$] $a_i \leftrightarrow b_i$,
\item[${\rm B_1}(a_3,b_1,b_2,c_1)$] $a_i \rightarrow b_i$, $b_i \rightarrow -c_i$, $c_i \rightarrow a_i$,
\item[${\rm B_1}(a_1,b_3,c_1,c_2)$] $a_i \rightarrow c_i$, $b_i \rightarrow a_i$, $c_i \rightarrow b_i$,
\item[${\rm B_1}(a_3,b_1,c_1,c_2)$] $a_i \leftrightarrow c_i$, $b_i \rightarrow -b_i$, 
\end{description}
from ${\rm B_1}(a_1,a_2,b_1,c_3)$, 
\item[2) ${\rm B_1}(a_2,a_3,b_1,c_3)$]
\begin{eqnarray}
\tilde{m}_D=\pm\sqrt{M_{ee}}
\left(
  \begin{array}{ccc}
     \frac{f_{e\mu}^{\rm B1}}{f_{\mu\tau}^{\rm B1}}f^{\rm B1} & 0 & -\frac{i\sigma f_{e\mu}^{\rm B1}\sqrt{f_{\tau\tau}^{\rm B1}}}{f_{\mu\tau}^{\rm B1}}   \\
    0  &  \frac{f_{\mu\tau}^{\rm B1}}{\sqrt{f_{\tau\tau}^{\rm B1}}} & \frac{i\sigma f_{\mu\tau}^{\rm B1}}{\sqrt{f_{\tau\tau}^{\rm B1}}} \\
   0 & \sqrt{f_{\tau\tau}^{\rm B1}} & 0 \\
  \end{array}
\right),
\end{eqnarray}
and
\begin{description}
\item[${\rm B_1}(a_2,a_3,b_3,c_1)$] $b_i \rightarrow c_i$, $c_i \rightarrow -b_i$,
\item[${\rm B_1}(a_1,b_2,b_3,c_3)$] $a_i \leftrightarrow b_i$,
\item[${\rm B_1}(a_3,b_2,b_3,c_1)$] $a_i \rightarrow b_i$, $b_i \rightarrow c_i$, $c_i \rightarrow a_i$,
\item[${\rm B_1}(a_1,b_3,c_2,c_3)$] $a_i \rightarrow c_i$, $b_i \rightarrow a_i$, $c_i \rightarrow b_i$,
\item[${\rm B_1}(a_3,b_1,c_2,c_3)$] $a_i \rightarrow c_i$, $c_i \rightarrow -a_i$,
\end{description}
from ${\rm B_1}(a_2,a_3,b_1,c_3)$,
\item[3) ${\rm B_1}(a_1,a_3,b_1,c_3)$]
\begin{eqnarray}
\tilde{m}_D=\pm\sqrt{M_{ee}}
\left(
  \begin{array}{ccc}
    0 & 0 & 1 \\
    \frac{i\sigma  f_{e\mu}^{\rm B1}}{\sqrt{f_{\tau\tau}^{\rm B1}}}f^{\rm B1} & \frac{f_{\mu\tau}^{\rm B1}}{\sqrt{f_{\tau\tau}^{\rm B1}}}  &  f_{e\mu}^{\rm B1}\\
   0 & \sqrt{f_{\tau\tau}^{\rm B1}} & 0 \\
  \end{array}
\right),
\end{eqnarray}
and
\begin{description}
\item[${\rm B_1}(a_1,a_3,b_3,c_1)$] $b_i \leftrightarrow c_i$,
\item[${\rm B_1}(a_1,b_1,b_3,c_3)$] $a_i \leftrightarrow b_i$,
\item[${\rm B_1}(a_1,b_3,c_1,c_3)$] $a_i \rightarrow c_i$, $b_i \rightarrow a_i$, $c_i \rightarrow b_i$,
\item[${\rm B_1}(a_3,b_1,b_3,c_1)$] $a_i \rightarrow b_i$, $b_i \rightarrow c_i$, $c_i \rightarrow a_i$,
\item[${\rm B_1}(a_3,b_1,c_1,c_3)$] $a_i \leftrightarrow c_i$,
\end{description}
from ${\rm B_1}(a_1,a_3,b_1,c_3)$.
\end{description}
For the ${\rm B_2}$ texture with $f^{\rm B2} =\sqrt{f_{\mu\mu}^{\rm B2}+\left( \frac{f_{\mu\tau}^{\rm B2}}{f_{e\tau}^{\rm B2}}\right)^2}$, there are three variations:
\begin{description}
\item[1) ${\rm B_2}(a_1,a_3,b_1,c_2)$]
\begin{eqnarray}
\tilde{m}_D=\pm\sqrt{M_{ee}}
\left(
  \begin{array}{ccc}
     0 & 0 & 1 \\
     f^{\rm B2} & \frac{i\sigma f_{\mu\tau}^{\rm B2}}{f_{e\tau}^{\rm B2}} & 0 \\
   0 & -i \sigma  f_{e\tau}^{\rm B2} & f_{e\tau}^{\rm B2} \\
  \end{array}
\right),
\end{eqnarray}
and
\begin{description}
\item[${\rm B_2}(a_1,a_3,b_2,c_1)$] $b_i \rightarrow -c_i$, $c_i \rightarrow b_i$,
\item[${\rm B_2}(a_1,b_1,b_3,c_2)$] $a_i \leftrightarrow b_i$,
\item[${\rm B_2}(a_2,b_1,b_3,c_1)$] $a_i \rightarrow b_i$, $b_i \rightarrow -c_i$, $c_i \rightarrow a_i$,
\item[${\rm B_2}(a_1,b_2,c_1,c_3)$] $a_i \rightarrow c_i$, $b_i \rightarrow a_i$, $c_i \rightarrow b_i$,
\item[${\rm B_2}(a_2,b_1,c_1,c_3)$] $a_i \leftrightarrow c_i$, $b_i \rightarrow -b_i$,
\end{description}
from ${\rm B_2}(a_1,a_3,b_1,c_2)$,
\item[2) ${\rm B_2}(a_2,a_3,b_1,c_2)$]
\begin{eqnarray}
\tilde{m}_D=\pm\sqrt{M_{ee}}
\left(
  \begin{array}{ccc}
     \frac{f_{e\tau}^{\rm B2}}{f_{\mu\tau}^{\rm B2}} f^{\rm B2} & 0 &  -\frac{i\sigma f_{e\tau}^{\rm B2}\sqrt{f_{\mu\mu}^{\rm B2}}}{\sqrt{f_{\mu\tau}^{\rm B2}}} \\
     0 & \sqrt{f_{\mu\mu}^{\rm B2}} & 0 \\
   0 & \frac{f_{\mu\tau}^{\rm B2}}{\sqrt{f_{\mu\mu}^{\rm B2}}} & \frac{i\sigma f_{\mu\tau}^{\rm B2}}{\sqrt{f_{\mu\mu}^{\rm B2}}} \\
  \end{array}
\right),
\end{eqnarray}
and
\begin{description}
\item[${\rm B_2}(a_2,a_3,b_2,c_1)$] $b_i \rightarrow c_i$, $c_i \rightarrow -b_i$,
\item[${\rm B_2}(a_1,b_2,b_3,c_2)$] $a_i \leftrightarrow b_i$,
\item[${\rm B_2}(a_2,b_2,b_3,c_1)$] $a_i \rightarrow b_i$, $b_i \rightarrow c_i$, $c_i \rightarrow -a_i$,
\item[${\rm B_2}(a_1,b_2,c_2,c_3)$] $a_i \rightarrow c_i$, $b_i \rightarrow a_i$, $c_i \rightarrow b_i$,
\item[${\rm B_2}(a_2,b_1,c_2,c_3)$] $a_i \rightarrow c_i$, $c_i \rightarrow -a_i$,
\end{description}
from ${\rm B_2}(a_2,a_3,b_1,c_2)$,
\item[3) ${\rm B_2}(a_1,a_2,b_1,c_2)$]
\begin{eqnarray}
\tilde{m}_D=\pm\sqrt{M_{ee}}
\left(
  \begin{array}{ccc}
    0 & 0 & 1 \\
    0 & \sqrt{f_{\mu\mu}^{\rm B2}} & 0  \\
    \frac{i \sigma f_{e\tau}^{\rm B2}}{\sqrt{f_{\mu\mu}^{\rm B2}}} f^{\rm B2} & \frac{f_{\mu\tau}^{\rm B2}}{\sqrt{f_{\mu\mu}^{\rm B2}}} & f_{e\tau}^{\rm B2}\\
  \end{array}
\right),
\end{eqnarray}
and
\begin{description}
\item[${\rm B_2}(a_1,a_2,b_2,c_1)$] $b_i \leftrightarrow c_i$,
\item[${\rm B_2}(a_1,b_1,b_2,c_2)$] $a_i \leftrightarrow b_i$,
\item[${\rm B_2}(a_1,b_2,c_1,c_2)$] $a_i \rightarrow c_i$, $b_i \rightarrow a_i$, $c_i \rightarrow b_i$,
\item[${\rm B_2}(a_2,b_1,b_2,c_1)$] $a_i \rightarrow b_i$, $b_i \rightarrow c_i$, $c_i \rightarrow a_i$,
\item[${\rm B_2}(a_2,b_1,c_1,c_2)$] $a_i \leftrightarrow c_i$,
\end{description}
from ${\rm B_2}(a_1,a_2,b_1,c_2)$.
\end{description}
For the ${\rm B_3}$ texture with $f^{\rm B3} =\frac{f_{\mu\tau}^{\rm B3}}{\sqrt{f_{\tau\tau}^{\rm B3}-(f_{e\tau}^{\rm B3})^2}}$, there is only one variation:
\begin{description}
\item[${\rm B_3}(a_2,b_1,b_3,c_1)$]
\begin{eqnarray}
\tilde{m}_D=\pm\sqrt{M_{ee}}
\left(
  \begin{array}{ccc}
     1 & 0 & 0\\
     0 & -i\sigma f^{\rm B3} & f^{\rm B3} \\
     f_{e\tau}^{\rm B3} & 0 & \frac{f_{\mu\tau}^{\rm B3}}{f^{\rm B3}} \\
  \end{array}
\right),
\end{eqnarray}
and
\begin{description}
\item[${\rm B_3}(a_2,b_1,c_1,c_3)$] $b_i \rightarrow -c_i$, $c_i \rightarrow b_i$,
\item[${\rm B_3}(a_1,a_3,b_2,c_1)$] $a_i \leftrightarrow b_i$,
\item[${\rm B_3}(a_1,b_2,c_1,c_3)$] $a_i \rightarrow b_i$, $b_i \rightarrow -c_i$, $c_i \rightarrow a_i$,
\item[${\rm B_3}(a_1,b_1,b_3,c_2)$] $a_i \leftrightarrow c_i$, $b_i \rightarrow -b_i$
\item[${\rm B_3}(a_1,a_3,b_1,c_2)$] $a_i \rightarrow c_i$, $b_i \rightarrow a_i$, $c_i \rightarrow b_i$,
\end{description}
from ${\rm B_3}(a_2,b_1,b_3,c_1)$.
\end{description}
For the ${\rm B_4}$ texture with $f^{\rm B4} =\frac{f_{\mu\tau}^{\rm B4}}{\sqrt{f_{\mu\mu}^{\rm B4}-(f_{e\mu}^{\rm B4})^2}}$, there is only one variation:
\begin{description}
\item[${\rm B_4}(a_3,b_1,b_2,c_1)$]
\begin{eqnarray}
\tilde{m}_D=\pm\sqrt{M_{ee}}
\left(
  \begin{array}{ccc}
     1 & 0 & 0\\
     f_{e\mu}^{\rm B4} & 0 & \frac{f_{\mu\tau}^{\rm B4}}{f^{\rm B4}} \\
     0 & -i\sigma f^{\rm B4} & f^{\rm B4}\\
  \end{array}
\right),
\end{eqnarray}
and
\begin{description}
\item[${\rm B_4}(a_3,b_1,c_1,c_2)$] $b_i \rightarrow -c_i$, $c_i \rightarrow b_i$,
\item[${\rm B_4}(a_1,a_2,b_3,c_1)$] $a_i \leftrightarrow b_i$,
\item[${\rm B_4}(a_1,b_3,c_1,c_2)$] $a_i \rightarrow b_i$, $b_i \rightarrow -c_i$, $c_i \rightarrow a_i$,
\item[${\rm B_4}(a_1,b_1,b_2,c_3)$] $a_i \leftrightarrow c_i$, $b_i \rightarrow -b_i$,
\item[${\rm B_4}(a_1,a_2,b_1,c_3)$] $a_i \rightarrow c_i$, $b_i \rightarrow a_i$, $c_i \rightarrow b_i$,
\end{description}
from ${\rm B_4}(a_3,b_1,b_2,c_1)$.
\end{description}
%

%%----------------------------------------------------------------------------------
\section{\label{sec:summary}Summary and discussions}
%%----------------------------------------------------------------------------------
If we require a nonvanishing Majorana effective mass $M_{ee}$ for the neutrinoless double beta decay, only four textures are compatible with observed data in the two zero flavor neutrino mass matrix scheme. We have obtained the complete list of the possible textures of the four zero Dirac neutrino mass matrix $m_D$ compatible with these four flavor neutrino mass matrix textures. We show that the four zero Dirac neutrino mass matrices are proportional to $\sqrt{M_{ee}}$ by explicit analytical analysis.

Finally, we would like to show the usefulness of the explicit analytical expressions of $m_D \propto\sqrt{M_{ee}}$. Since the rescaled Dirac mass matrix with $\sqrt{M_{ee}}$ is known, it is interesting to apply the present method to the creation of the baryon number of the universe via the leptogenesis, which depends on the details of the Dirac neutrino masses.

The baryon number of the universe in the leptogenesis scenario depends on the decay of the lightest right-handed neutrinos \cite{leptogenesis1,leptogenesis2,leptogenesis3,leptogenesis4,leptogenesis5,leptogenesis6,FlavoredLeptogenesis1,FlavoredLeptogenesis2,FlavoredLeptogenesis3,FlavoredLeptogenesis4}, where we assume that $M_1 \ll M_2, M_3$.  The CP asymmetry parameter from the decay of the lightest right-handed neutrino $N_1$ is obtained from 
\begin{eqnarray}
\epsilon = \sum_i \epsilon_i= -\frac{3M_1}{16\pi v^2}\sum_i \tilde{\epsilon}_i,
\end{eqnarray}
where $i=e,\mu,\tau=1,2,3$, $v \simeq 174$ GeV and 
\begin{eqnarray}
\tilde{\epsilon}_i &=& \frac{I_i^{12} + I_i^{13}}{\sum_{i=1}^3 \vert a_i \vert^2},
\label{Eq:tilde_epsilon_alpha}
\end{eqnarray}
with
\begin{eqnarray}
I_i^{12} &=& {\rm Im}[a_i^\ast b_i(a_1^\ast b_1 + a_2^\ast b_2 + a_3^\ast b_3)], \nonumber \\
I_i^{13} &=& {\rm Im}[a_i^\ast c_i(a_1^\ast c_1 + a_2^\ast c_2 + a_3^\ast c_3)] 
\label{Eq:I_alpha}
\end{eqnarray}
The baryon number in the co-moving volume is calculated to be
\begin{eqnarray}
Y_B &\simeq& -0.01 \epsilon \eta\left(\sum_{i=1}^3 |a_i|^2 \right),
\label{Eq:YL}
\end{eqnarray}
for $10^{12} {\rm GeV} \le M_1$ (one flavor dominant case \cite{FlavoredLeptogenesis1,FlavoredLeptogenesis2,FlavoredLeptogenesis3,FlavoredLeptogenesis4}) where washout effect on $\epsilon_i$ in the expanding universe is controlled by  
\begin{eqnarray}
\eta(x) = \left( \frac{8.25\times 10^{-3} {\rm eV}}{x} + \left( \frac{x}{2\times 10^{-4} {\rm eV}} \right)^{1.16} \right)^{-1}.
\end{eqnarray}
The baryon-photon ratio $\eta_B$ is estimated to be $\eta_B=7.04Y_B$. 

%--------------------------------------------------------------------
\begin{figure}[t]
\begin{center}
\includegraphics{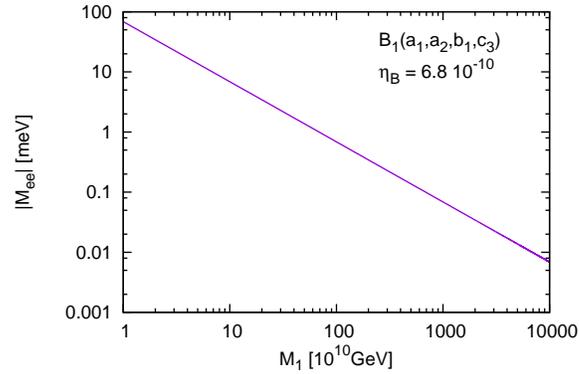}
\caption{$|M_{ee}|$ - $M_1$ plane for $\eta_B=6.8\times 10^{-10}$.}
\label{fig:B1}
\end{center}
\end{figure}
%--------------------------------------------------------------------
Since the relation of $m_D \propto \sqrt{M_{ee}}$ is obtained, we predict that $Y_B \propto |M_{ee}|$. To evaluate $Y_B$, we choose the ${\rm B_1}(a_1,a_2,b_1,c_3)$ texture as an illustration. In the one flavor dominant case, we obtain 
\begin{eqnarray}
Y_B = \frac{0.003M_1|M_{ee}|}{16\pi v^2} \frac{{\rm Im}\left[f_{\tau\tau}^{\rm B1 *} \left(\frac{f_{\mu\tau}^{\rm B1}}{f_{e\mu}^{\rm B1}}\right)^2\right]}{\left| f_{\tau\tau}^{\rm B1} + \left(\frac{f_{\mu\tau}^{\rm B1}}{f_{e\mu}^{\rm B1}}\right)^2 \right|} \eta,
\end{eqnarray}
for $M_1 \ll M_2 \le M_3$. The baryon asymmetry via leptogenesis scenario with four zero Dirac neutrino mass matrix has already been extensively studied in the literature \cite{fourZeroDirac1,fourZeroDirac2,fourZeroDirac3,fourZeroDirac4,fourZeroDirac5,fourZeroDirac6,fourZeroDirac7} and the relation of $Y_B \propto |M_{ee}|$ is obtained by numerical calculations; however, we show the relation of $Y_B \propto |M_{ee}|$ by exact analytical expressions. It is our advantage to use the explicit analytical expressions of $m_D$, which will be complementary to previously reported numerical analyses.

Although, we have reached our main goal of our discussions to see the usefulness of the explicit analytical expressions, a numerical calculation may be required to confirm results of our discussions. We use $\sin^2 \theta_{12} = 0.304$, $\sin^2 \theta_{23} = 0.452$ and $\sin^2 \theta_{13} = 0.0219$ for three mixing angles and $\delta=270^\circ$ for the CP-violating Dirac phase \cite{Nudata,T2K2015PRD}.  The octant of $\theta_{23}$ (i.e. lower octant $\theta_{23}<45^\circ$ or upper octant $\theta_{23}>45^\circ$) is still unresolved problem. According to Dev et.al. \cite{twoZeroFlavor2}, we take lower octant of $\theta_{23}$ for the $B_1$ texture with the so-called normal ordering of the neutrino mass eigenstates. The theoretically expected half-life of the neutrinoless double beta decay is proportional to the effective Majorana neutrino mass $\vert M_{ee}\vert$. 

Fig. \ref{fig:B1} shows the prediction for $|M_{ee}|$ as a function of $M_1$ for ${\rm B_1}(a_1,a_2,b_1,c_3)$ texture. For example, we obtain $M_1\sim 10^{12}$ GeV and $|M_{ee}| \sim 1$ meV for $\eta_B=6.8\times 10^{-10}$. The estimated magnitude of the effective Majorana mass from the experiments is $\vert M_{ee} \vert \lesssim 0.20-2.5$ eV \cite{doubleBetaDecay}. In the future experiments, more sensitivity $\vert M_{ee} \vert \simeq$ a few meV will be reached \cite{doubleBetaDecay}. We use one flavor dominant approximation \cite{FlavoredLeptogenesis1,FlavoredLeptogenesis2,FlavoredLeptogenesis3,FlavoredLeptogenesis4} for $M_1 = 10^{10} - 10^{14}$ GeV for the sake of simplicity. More general analysis will be found in our future study.

%%\appendix
%%-------------------------------------------------
%% Appendix
%%-------------------------------------------------

%%--------------------------------
%%References
%%--------------------------------
%\begin{thebibliography}{000} %for 3 digits
%\begin{thebibliography}{00}  %for 2 digits

\end{document}